%% file: ProvWiz.tex
\documentclass[fleqn,usenatbib]{mnras}

\usepackage[T1]{fontenc}
\usepackage{newtxtext, newtxmath}

\usepackage{graphicx}	% Including figure files
\usepackage{rotating}
\usepackage{multirow}
\usepackage{tikz}
\usepackage{subcaption}

\usepackage{fancyhdr}
\newcommand{\indep}{\perp \!\!\! \perp}

\title[Causal Estimation for UX Interventions]{Causal Effect Estimation for UX Interventions without Randomized Controlled Experiments}

\author[S. Juhl et al.]{
Sebastian Juhl,$^{1,2}$\thanks{E-mail: sebastian.juhl@sap.com}
Yi-Jun Li,$^{3}$
Johannes Osterhoff,$^{1}$
Dahye Park$^{4}$ and
Vasco Alexander Schmidt$^{1}$
\\
$^{1}$SAP SE, Walldorf, Germany\\
$^{2}$University of Missouri, Columbia, USA\\
$^{3}$SAP Co., Ltd., Xi'an, China\\
$^{4}$SAP Labs Korea, Seoul, South Korea
}

\date{Version: \today}

\pubyear{2026}

\begin{document}
\label{firstpage}
\pagerange{\pageref{firstpage}--\pageref{lastpage}}
\maketitle

\begin{abstract}
Evaluating the causal impact of user experience (UX) interventions in business software requires rigorous statistical methods, yet randomized controlled trials are often infeasible due to technical constraints, organizational limitations, or concerns about user equity. This study demonstrates how regression discontinuity (RD) design, a quasi-experimental approach, enables credible causal inference when randomized controlled experiments are unavailable. The RD design exploits natural discontinuities in treatment assignment (e.g., product deployment dates) to approximate randomization at a threshold, thereby identifying local average treatment effects without requiring explicit randomization or strong assumptions about confounders. We demonstrate the utility of this approach using a case study of a UX redesign in SAP HANA Cloud Central's Provisioning Wizard. Using the RD approach, we find that the redesigned wizard significantly increased task completion rates and reduced time-to-task completion. Placebo cutoff analyses confirm the validity of our causal estimates. This study contributes a practical methodological framework for evidence-based product evaluation in production environments where randomized experiments are infeasible, enabling organizations to rigorously quantify the value of design investments.
\end{abstract}


% remove header from first page
\renewcommand{\headrulewidth}{0pt}
\pagestyle{fancy}
\fancyhead{}
\fancyhead[LE]{\Large \thepage\hspace{.5cm} \textit{S. Juhl et al.}}
\fancyhead[RO]{\Large \textit{Causal Estimation for UX Interventions} \hspace{.5cm} \thepage}
\fancyfoot{}

\thispagestyle{empty}

%%%%%%%%%%%%%%%%%%%%%%%%%%%%%%%%%%%%%%%%%%%%%%%%%%
%%%%%%%%%%%%%%%%% BODY OF PAPER %%%%%%%%%%%%%%%%%%
\section{Introduction}
User experience (UX) improvements in business software represent substantial investments that organizations hope will enhance key performance indicators (KPIs) such as task completion rates, efficiency, and user satisfaction. However, quantifying the causal impact of design interventions remains challenging. While randomized controlled trials represent the gold standard for causal inference in UX research \citep[e.g.,][]{larsen2024, quin2024, Kohavi2017, kohavi2009}, they are frequently infeasible in production environments due to technical infrastructure limitations, organizational constraints, or concerns about inequitable user experiences. Consequently, many organizations must rely on observational data to evaluate whether design changes meaningfully improve user outcomes or merely correlate with temporal trends and confounding factors.

This tension between the need for rigorous causal inference and the practical constraints of production environments creates a critical gap in applied UX research methodology. While qualitative methods and descriptive analytics can document changes in user behavior, they cannot establish causality \citep[e.g.,][]{pearl2009}. Conversely, randomized experiments are often unavailable. This paper addresses this methodological challenge by demonstrating how regression discontinuity (RD) design, a quasi-experimental approach, can credibly estimate causal effects of UX improvements when randomized controlled experiments are infeasible \citep{thistlethwaite1960}.

The RD design is particularly well-suited to UX evaluation contexts because product redesigns inherently create sharp temporal thresholds. Users accessing a product before a deployment date receive the legacy interface, while those accessing afterward experience the redesigned version. This naturally occurring discontinuity in treatment assignment mirrors the structure of randomized experiments at the threshold, enabling causal inference without explicit randomization. Unlike alternative quasi-experimental methods that require strong and often untestable assumptions about observed confounders, the RD design imposes relatively mild assumptions, focusing instead on the comparability of users arbitrarily close to the deployment cutoff \citep[e.g.,][]{lee2010, hahn2001}.

To demonstrate the value of this approach for practitioners, we present a detailed case study of a UX redesign regarding the Provisioning Wizard in SAP HANA Cloud Central. The Provisioning Wizard is a critical component of HANA Cloud as improving its usability has direct implications for key business KPIs like the number of provisioned instances and operational efficiency. Despite the importance of this feature, the organization lacked a controlled experimentation infrastructure at the time of the redesign, necessitating an alternative evaluation strategy. By leveraging the deployment date as treatment assignment mechanism, we employ an RD design to estimate the causal effect of the redesigned wizard on two key outcomes: (1) task completion and (2) time to task completion.

This study contributes to the methodological toolkit of UX researchers and product managers in three ways. First, we provide a rigorous theoretical foundation for the RD approach, grounding it in the potential outcomes framework and causal inference theory. Second, we demonstrate practical implementation details, including bandwidth selection, robustness checks via placebo cutoffs, and interpretation of local average treatment effects (LATE) through stochastic simulations. Third, our empirical case demonstrates that the RD design can quantify the effects of design interventions when randomized controlled experiments are unavailable, thereby supporting evidence-based product decisions under real-world constraints.

%The remainder of this paper is structured as follows. After introducing the redesign of the Provisioning Wizard in HANA Cloud Central as our use case, we present the research strategy with an emphasis on the potential outcomes framework and RD designs. We then describe our data, operationalization, and empirical results. Finally, we discuss the implications of our findings for UX research practice and identify directions for future work.

\section{Use Case: Redesigning the Provisioning Wizard in SAP HANA Cloud Central}
SAP HANA Cloud Central is the main administrative tool for SAP HANA Cloud and is used to create, monitor, and administer SAP HANA Cloud instances. The Provisioning Wizard is a step-by-step UI that guides users through configuring and creating a new instance. While the wizard covers a wide range of configuration options, this comprehensiveness can slow users down and reduce completion rates, directly impacting the number of successfully provisioned instances. To address this, we analyzed usage data collected from March 2024 to March 2025 through Azure Monitor Application Insights and used the findings to improve the design in two phases.

\begin{figure}
\centering
\includegraphics[width=.4\textwidth]{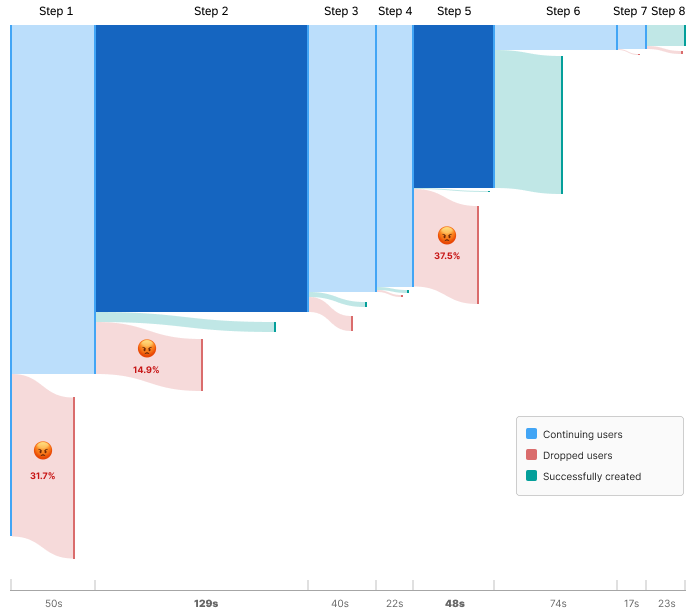}
\caption{Sankey diagram of the old provisioning wizard user flow. Step width represents average completion time and step height represents the proportion of users who continued or abandoned the wizard at each step. Highlighted steps (Step 2, Step 5) indicate the longest duration and highest drop-out rate.} \label{fig:sankey}
\end{figure}

As Figure~\ref{fig:sankey} illustrates, our descriptive analysis revealed two compounding issues. First, the time users spend in the wizard was concentrated in specific steps: on average, users spent the most time in Step 2 (129 seconds), where mandatory fields are concentrated. Step 5 (48 seconds), which surfaces optional advanced settings, took also notably longer on average than the remaining steps. Second, the drop-out rate reinforced this pattern: among the configuration steps, the highest abandonment occurred at Step 5 with 37.5\% and Step 2 with 14.9\% of all users who reached the respective step. Again, these two steps were the ones with the highest configuration demand. Together, we saw an overall wizard completion rate of only 39.5\%.

\begin{figure}
\centering
\begin{subfigure}{0.49\linewidth}
  \centering
  \includegraphics[width=\linewidth]{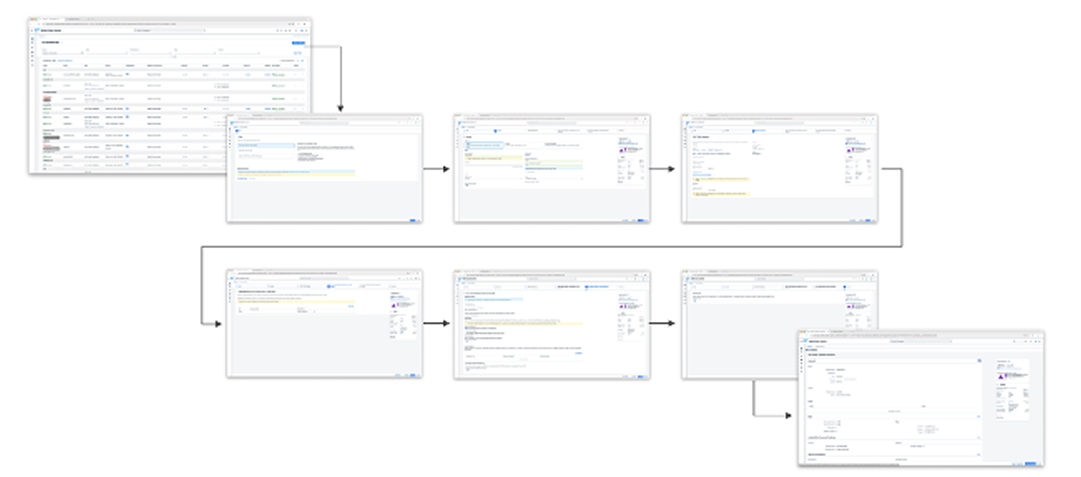}
  \caption{Old version: multi-step manual configuration}
  \label{fig:old}
\end{subfigure}
\hfill
\begin{subfigure}{0.49\linewidth}
  \centering
  \includegraphics[width=\linewidth]{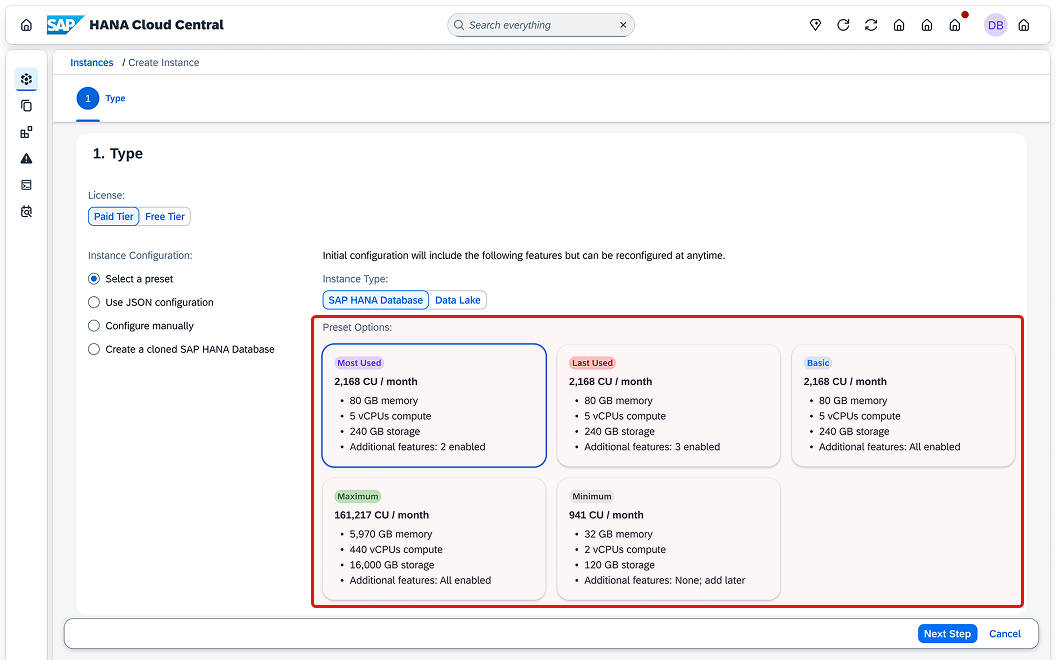}
  \caption{New version: Preset-based configuration}
  \label{fig:new}
\end{subfigure}
\caption{Comparison of the old and new version of the provisioning wizard. The old version requires six sequential manual configuration steps whereas the new version of the wizard introduces preset configurations (highlighted), allowing users to skip the step-by-step setup by selecting a predefined instance profile.}
\label{fig:both}
\end{figure}

Further analyses pointed to the opportunity for a simplified configuration path. New users had a notably longer average time to complete their task of creating a new instance than returning users (423 seconds vs. 356 seconds), suggesting the initial experience presents a particular challenge. Production landscape data reinforced this: 77.6\% of running instances had advanced options disabled and 34\% used entirely default configurations. Based on the usage data, we concluded that most users do not require deep customization but rather a quick, clear path to a working instance. Consequently, our aim was to offer a simplified alternative to the multi-step configuration flow displayed in Figure~\ref{fig:old}.

Based on these observations, we redesigned the Provisioning Wizard in two phases. In Phase 1, we introduced four configuration paths, shown in Figure~\ref{fig:new}: presets for users who want a quick start, manual configuration for users who want to have full control over the settings, clone for the replication of an existing instance, and JSON import for users with configuration files ready to use. The preset path directly addresses the friction identified in the analysis: users complete the provisioning of an instance in a single step by filling in only the essential fields, bypassing the configuration-heavy steps where delays and abandonment were most concentrated.

In Phase 2, we focused on improving the full configuration experience. Referring back to the step duration and drop-out rate data, we reorganized settings by priority: related options were grouped together, frequently adjusted or strategically important items were surfaced prominently, and rarely changed items were pushed to an advanced step. We also rearranged steps to reduce transitions and integrated inline help to allow users to complete the configuration without consulting external documentation. These improvements were made to reduce the time users need to complete the provisioning process and to increase the overall task completion rate.
 
\section{Research Strategy}
Evaluating the causal impact of UX improvements constitutes a challenge since it requires specific statistical techniques and a carefully crafted research design \citep{pearl2009}. Determining whether observed changes in user behavior are causally attributable to a design intervention or merely reflect confounding factors and temporal trends. Randomized controlled trials represent the gold standard for causal inference, as randomization eliminates confounding. However, randomized controlled experiments are not always feasible in production environments due to technical constraints, organizational resistance, or ethical considerations regarding user experience equity \citep{kohavi2009}.

In such contexts, quasi-experimental designs can provide credible causal estimates by leveraging natural variation in the timing or implementation of interventions. In this study, we utilize a sharp regression discontinuity (RD) design, a quasi-experimental method that exploits a threshold determining exposure to treatment. We motivate this approach using the potential outcomes framework, which provides a rigorous mathematical foundation for causal inference.

\subsection{The Potential Outcomes Framework}
Following the Rubin Causal Model\footnote{This model is also known as the Neyman-Rubin Causal Model because the potential outcomes framework was first used by \citet{neyman1923} in the context of randomized experiments. \citet{rubin1974} later extended the framework to observational (nonrandomized) studies.} \citep{rubin1974}, we define potential outcomes as the outcomes that \textit{would} occur under each treatment state \citep[see also][]{larsen2024, rubin2005, angrist1996}. For each user $i$, let $T_i \in \{0, 1\}$ be the binary treatment state\footnote{For the sake of simplicity, this discussion pertains to the case of a binary treatment state where users get selected into a treatment and a control group. All results hold for the case of more than two treatment states as well.} and $Y_i(T)$ the potential outcome for user $i$ under treatment regime $t$. Hence, $Y_i(1)$ is the outcome if user $i$ receives the treatment – the revised Provisioning Wizard – and $Y_i(0)$ the outcome if user $i$ is in the control group that does not receive the treatment and instead sees the old interface. The causal impact of any treatment is simply the difference in the potential outcomes for each user:

\begin{equation}\label{eq:causalimpcat}
Y_i(1) - Y_i(0)\hspace{.5cm} \forall i.
\end{equation}

However, the ``Fundamental Problem of Causal Inference'' \citep[947]{holland1986} is that, for each user, we only observe one of these potential outcomes shown in Equation \ref{eq:causalimpcat}. Each user only sees the revised or the old Provisioning Wizard. Hence, depending on the treatment state $T$, we either observe $Y_i(1)$ or $Y_i(0)$. The unobserved counterfactual – what would have happened under the alternative condition – is never observed. Therefore, we cannot compute the unit-level causal effect of a treatment because this would require us to be able to ``unwind time and go back and expose the $i$th unit to the other treatment'' \citep[323]{rubin2005}.

This does not imply that we cannot make any causal claims at all. Instead of attempting to estimate the unit-level causal effect, there is a statistical solution to the Fundamental Problem of Causal Inference that focuses on a different quantity of interest, called the \textit{estimand} in statistics. The idea is to estimate the average causal effect of a treatment within a pre-defined population of interest $U$, e.g., all users of a specific software product, where $i \in U \forall i$ \citep{angrist1996, holland1986}. Following this approach, the average treatment effect (ATE) on the population of interest is defined as:

\begin{equation}\label{eq:ATE}
\begin{array}{rl}
\text{ATE} &= \mathbb{E}[Y_i(1) - Y_i(0)]\\
&= \mathbb{E}[Y_i(1)] - \mathbb{E}[Y_i(0)],
\end{array}
\end{equation}

where $\mathbb{E}[\cdot]$ is the expectation operator. In contrast to Equation \ref{eq:causalimpcat} which is inherently impossible to observe, the estimand in Equation \ref{eq:ATE}, the $ATE$ over the user population $U$, can be estimated \citep[e.g.,][]{holland1986, rubin1974}. \citet{neyman1923} shows that the difference in the observed means between the treatment and control groups is an unbiased estimator of the ATE \citep[see also][324]{rubin2005}:

\begin{equation}\label{eq:ATE_hat}
\begin{array}{rl}
\widehat{ATE} &= \bar{Y}(1) - \bar{Y}(0)\\
&= \frac{1}{n_1}\sum_{i:T_i=1} Y_i - \frac{1}{n_0}\sum_{i:T_i=0} Y_i,
\end{array}
\end{equation}

where $n_1$ is the number of users in the treatment group ($T = 1$) and $n_0$ is the number of users in the control group ($T = 0$) \citep[137]{larsen2024}.

At this point, it is important to consider the mechanism which is the procedure determining the treatment state $T \in \{0, 1\}$ for each user \citep{rubin2005}. In randomized controlled experiments, randomization ensures that the treatment status is independent of the realized outcome $Y$ as well as any other observed or unobserved factor $Z$ such that $T_i \indep (Y_i, Z_i)$. Consequently, the probability that user $i$ receives the treatment, the redesigned Provisioning Wizard, is independent of $Y_i$ and $Z_i$ which means that $\Pr(T_i = 1 | Y_i, Z_i) = \Pr(T_i = 1)$. Under this condition, Equation \ref{eq:ATE_hat} is an unbiased estimator of the ATE since $\mathbb{E}[Y_i | T_i = 1] - \mathbb{E}[Y_i | T_i = 0] = \mathbb{E}[Y_i(1)] - \mathbb{E}[Y_i(0)] $.

In the absence of a fully controlled experimental setting with randomized treatment assignment, however, unobserved factors such as user sophistication, job role, or task urgency captured by $Z$ may affect both the treatment state $T$ and the potential outcomes $Y(1)$ and $Y(0)$. This introduces confounding bias since the true ATE now becomes $\mathbb{E}[Y_i(1) - Y_i(0) | Z_i]$ which does not equal $\mathbb{E}[Y_i(1) - Y_i(0)]$ unless $T_i \indep (Y_i, Z_i)$.

To obtain an unbiased estimate, we would need to condition on all confounding factors $Z_i$. However, since $Z$ may contain a myriad of factors – some of which are inherently unmeasurable – we cannot reliably condition on all relevant confounders. As a result, the group averages $\bar{Y}(1)$ and $\bar{Y}(0)$ suffer from omitted variable bias, and any causal inference based on their difference is unreliable.

Different statistical techniques have been proposed that use specific assumptions about the assignment mechanism to allow for causal statements in observational studies where a randomized treatment assignment is impossible \citep{rubin2005, holland1986}. In our use case, we implement a widely applied sharp RD design, introduced by \citet{thistlethwaite1960}.

%As with all statistical techniques, the validity of the inferences on the estimand hinges on certain assumptions, some of which are inherently untestable.
%``stable unit-treatment value assumption''(SUTVA) \citep[591]{rubin1980}

\subsection{Regression Discontinuity Design}
The RD design exploits a natural threshold in a \textit{running} or \textit{forcing} variable $R$ to approximate randomization in treatment assignment. \citep[e.g.,][]{imbens2008}. At a specific and known cutoff value $\tau$ of the running variable, the probability of receiving the treatment changes discontinuously. For sharp RD designs, the change is deterministic such that $T_i = 0 \forall R_i < \tau$ and $T_i = 1 \forall R_i \geq \tau$. When studying the effect of student scholarships on career aspirations, for example, \citet{thistlethwaite1960} exploit the fact that students need to achieve a certain test score to receive a scholarship. Individuals just below and just above the cutoff are likely to be comparable in observable and unobservable characteristics – but differ in treatment status as only students with scores above the cutoff receive the scholarship. In fact, this design can be frequently observed in many different occasions, leading to the broad application of RD designs across disciplines \citep{lee2010, hahn2001}.

In contrast to other quasi-experimental designs, the RD design imposes rather mild assumptions to be able to identify a causal effect. Besides the ``stable unit-treatment value assumption'' \citep[591]{rubin1980} that requires that the outcome for one user should be unaffected by the treatment assignment of any other user \citep[see also][]{cox1958}, RD designs impose a continuity assumption and a conditional independence assumption. While the continuity assumption requires that only the running variable $R$ (but no other variables) changes discontinuously at the cutoff $\tau$, the conditional independence assumption states that users do not self-select into treatment or control group \citep{hahn2001}.

Conditional on the validity of these assumptions, users arbitrarily close to $\tau$ are comparable and we can estimate a local average treatment effect (LATE) at the cutoff $\tau$. The estimator is given by

\begin{equation}
\widehat{LATE}(\tau) = \lim_{r \to \tau^+} \mathbb{E}[Y_i|R_i=r] - \lim_{r \to \tau^-} \mathbb{E}[Y_i|R_i=r],
\end{equation}

where $r$ is a realization of the running variable $R$ \citep[see also][825]{cattaneo2022}. This represents the causal effect at the threshold. Under the continuity assumption, we can run a local linear regression with a dummy variable indicating the treatment status using the data points in the neighborhood of $\tau$ to estimate the LATE \citep{lee2010}. An important implication of this \textit{local} identification strategy is that the LATE is, by construction, restricted to users in the vicinity of the cutoff $\tau$.

In the Provisioning Wizard redesign, the running variable $R$ is the timestamp at which user $i$ initiated the wizard. The treatment cutoff is set at date $\tau$: users who accessed the wizard before $\tau$ were exposed to the original interface (control), while users accessing the wizard after $\tau$ received the redesigned interface (treatment). Since the users cannot anticipate the redesign, self-selection into treatment is impossible.

\section{Data and Operationalization}
We analyze usage logs from SAP HANA Cloud Central productive landscapes from June and July, 2025. The data is at the event-level. For the analysis, we only consider data from productive landscapes and collect all events from accessing the Provisioning Wizard until the last action on the page. We aggregate the event-level data that records all individual click actions a user performs during one session so that the unit of analysis is on the operation-level which encompasses all actions a user performs after accessing the Provisioning Wizard. Importantly, the logs are only available for users that accept cookies. Furthermore, we restrict the analysis to the provisioning of databases. While users can use the Provisioning Wizard to create a database or a data lake, the type of instance affects both the completion rate and the average time users need to provision an instance. Since our data only contains $115$ operations in which users want to provision a data lake, we exclude these cases from the analysis.

The running variable $R_i$ is the date at which user $i$ initiated a provisioning session. The redesign was deployed at the beginning of July 2025, which represents the cutoff $\tau$.\footnote{The exact deployment date differ by landscape. Therefore, we specify different cutoff values for the $33$ landscapes in our sample and utilize information on which landscape the users operate to calculate the distance of their operations to the respective landscape-specific cutoff date.} Overall, the analysis encompasses $922$ unique users who accessed the Provisioning Wizard $2075$ times in the period from June 01, 2025 to July 31, 2025. The old wizard was used in $911$ of the cases and the new one in $1164$.

The analysis focuses on two outcome variables. First, \textit{task completion} is a binary indicator with $Y^c_i = 1$ if user $i$ completed the provisioning workflow and $Y^c_i = 0$ otherwise. Second, \textit{time to task completion} ($Y^t_i$) measures the seconds it takes user $i$ to successfully complete the workflow. Hence, we restrict the analysis in this case to the operations that led to the successful provisioning of a HANA database to evaluate if the redesigned wizard allows users to be more efficient.

\section{Results}
\subsection{Completion Rates and Completion Probabilities}
We first investigate if the redesigned Provisioning Wizard affects the likelihood that users successfully create a database instance. At the aggregate level, using all operations in our data, we find that the completion rate for the old wizard is $39.48$\% while the new wizard has a completion rate of $47.41$\%. To test whether the increase in the completion rate is statistically significant, we employ a two-sided $\chi^2$-test with Yates’ continuity correction. The test suggests that the null hypothesis of equality in proportions can be rejected (p-value = $0.001$).\footnote{Fisher's Exact test for count data yields similar results with a p-value of $0.001$.} Hence, we conclude that the new Provisioning Wizard increases the completion rate by $7.92$pp with a 95\% confidence interval (CI) within [$3.08$pp; $12.77$pp].

Since we have data at the operations-level, we can also calculate the change in the completion probabilities per operation. To this end, we start by comparing the completion probabilities before and after the rollout of the new wizard using a logistic regression model where the outcome is $Y^c$, the binary indicator of whether or not an operation has led to the successful provisioning of a database. The only explanatory variable in the model is a binary indicator of whether or not the new wizard has been used. The regression model reveals that, all else equal, the odds for successfully creating an instance increase with the new Provisioning Wizard by $38.16$\% [$13.87$\%; $67.72$\%].

Putting these results into perspective, we can simulate how many database instances would have been successfully provisioned would the new Provisioning Wizard already have been available in June 2025.\footnote{For the simulation, we use the coefficient estimates and the estimated variance-covariance matrix from the logistic regression model to set up a multivariate normal distribution from which we sample $1,000$ coefficient pairs.} From the total of $515$ Provisioning Wizard operations that did not lead to the successful provisioning of a database instance using the old wizard, our results suggest that $196$ [$78$; $351$] of them would have lead to a successful creation of a database instance with the redesigned wizard, all else equal.\footnote{The simulation relies on two simplifying assumptions: 1) Users who enter the Provisioning Wizard want to create a database or data lake instance. 2) Users who dropped off without creating an instance do not create the desired instance at a later point in time.}

After finding a significant correlation between the new Provisioning Wizard and the completion rate, we now seek to establish a causal relationship between these two variables using the RD design introduced above. The analysis consists of a local logistic regression with a binary indicator of whether or not the user uses the new wizard, the differences in days to the cutoff, and an interaction term between the two which allows the effect of the running variable to vary between the treated and untreated units. We use an empirical bandwidth selector to identify a mean squared error (MSE) optimal bandwidth that balances statistical power with the validity assumption that observations far from $\tau$ may not be comparable. Based on this procedure, we use a window of 12 days before and after the cutoff date \citep[see][]{calonico2015}.

Using the optimal window of 12 days prior and after the deployment of the redesigned wizard, the RD analysis here is based on $293$ observations for the old and $353$ for the new wizard. Table \ref{tab:rdd_bin} shows the results.

\begin{table} \centering 
  \caption{Coefficient Estimates from the Local Logistic Regression Model} 
  \label{tab:rdd_bin} 
\begin{tabular}{@{\extracolsep{5pt}}lc} 
\\[-1.8ex]\hline 
\hline \\[-1.8ex] 
 & \multicolumn{1}{c}{\textit{Dependent variable:}} \\ 
\cline{2-2} 
\\[-1.8ex] & Completed – $Y^c$\\ 
\hline \\[-1.8ex] 
 T & 0.738$^{**}$ \\ 
  & (0.319) \\ 
  & \\ 
 R & $-$0.007 \\ 
  & (0.035) \\ 
  & \\ 
 T $\times$ R & $-$0.046 \\ 
  & (0.046) \\ 
  & \\ 
 Constant & $-$0.630$^{**}$ \\ 
  & (0.268) \\ 
  & \\ 
\hline \\[-1.8ex] 
Observations & 646 \\ 
Log Likelihood & $-$433.539 \\ 
Akaike Inf. Crit. & 875.078 \\ 
\hline 
\hline \\[-1.8ex] 
\textit{Note:}  & \multicolumn{1}{r}{$^{*}$p$<$0.1; $^{**}$p$<$0.05; $^{***}$p$<$0.01} \\ 
\end{tabular} 
\end{table}

While Table \ref{tab:rdd_bin} already shows that the coefficient associated with the treatment state indicator $T$ is statistically significant, we simulate the LATE to ease the substantive interpretation of the results. The simulations reveal that the completion probability of each operation increases by $17.9$pp [$3.56$pp; $29.69$pp] with the redesigned Provisioning Wizard. With an empirical p-value of $0.007$, this effect is statistically significant.

To validate the findings, we repeat the analysis but instead of using the actual cutoff $\tau$ that determines treatment assignment, we specify a set of artificial placebo cutoff dates to determine treatment assignment. The intuition is that, since $\tau$ is the only point in $R$ where the outcome is expected to change discontinuously due to treatment exposure, using a different cutoff value should yield insignificant results \citep{cattaneo2022}. For our analysis, we specify placebo cutoffs that are $3$, $5$, $7$, and $10$ days prior and after $\tau$. Our placebo analysis confirms that none of the effects is statistically significant, providing additional evidence that the discontinuous change in the completion probability is caused by the redesign of the Provisioning Wizard.

Overall, we conclude that the new Provisioning Wizard increases the completion rate. Our results further indicate that there is a causal relationship between the two which provides empirical evidence for the claim that the redesigned Provisioning Wizard successfully increases the completion rate.

\subsection{Time to Task Completion}
Turning to the second metric of interest, we evaluate the effect of the redesigned Provisioning Wizard on the time users need to successfully create a HANA database instance. To this end, we restrict the sample to operations that have led to a successful creation of a HANA database instance. We do this in order to exclude users that simply want to explore the options available or that accidentally opened the Provisioning Wizard in HANA Cloud Central. Again, we start with a simple comparison of the average time and then move on to an RD design.

\begin{figure}
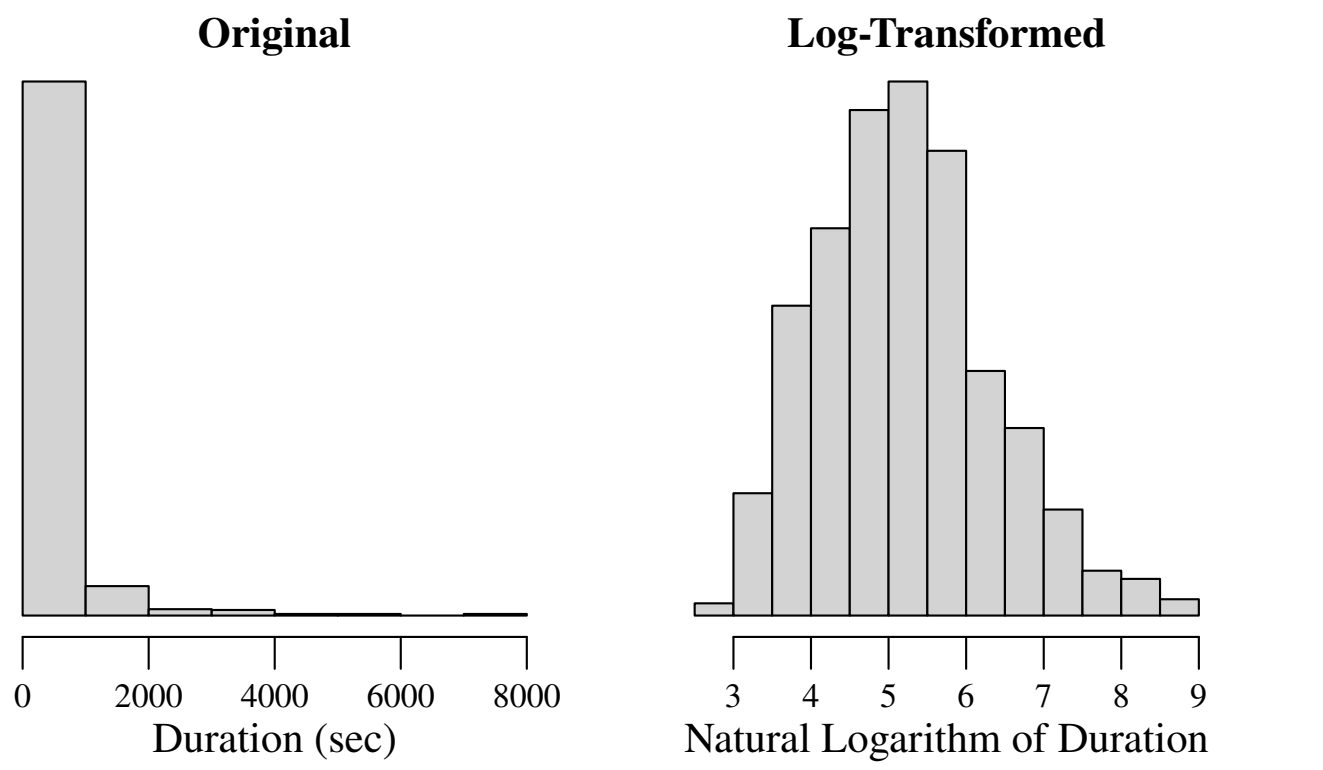

\include{distribution}
\caption{Frequency Distribution of $Y^t$ and $log(Y^t)$. The log transformation results in a distribution that resembles a normal distribution.}\label{fig:distribution}
\end{figure}

An additional difficulty is the skewed distribution of the variable $Y^t$ – the time it takes from entering the Provisioning Wizard to successfully create a database instance in seconds. While the quickest user needed $17.28$ seconds, it took the slowest user $7781.47$ seconds, more than 2 hours, to create a HANA Database instance. The arithmetic mean is $378.19$ seconds and the median $163.87$ seconds, indicating the skewness in the data. To make sure that the necessary assumptions we need to make for the following statistical tests are properly met, we apply a log transformation to the variable. As Figure \ref{fig:distribution} shows, this transformation makes the data appear normally distributed which is of crucial importance for the validity of the statistical tests we apply.

By transforming the results from the log-scale back to the original scale (duration in seconds), the average time users need to successfully create a database instance with the old wizard is $193.15$ seconds while with the redesigned one, it decreases to $172.27$ seconds. To test whether this difference of $20.88$ seconds is statistically significant, we implement a Welch's t-test of independent samples. The test suggests that, although the new wizard leads to a faster creation of an instance, the difference is not statistically significant at conventional confidence levels (p-value = $0.172$).

One reason for this may be that a new process requires users to learn. Although the new workflow may be more efficient, users need to adapt to a new, unfamiliar process which leads to an increase in the time users need to complete a specific task. To account for this, we utilize additional data from the logs and identify users who open the Provisioning Wizard for the first time. Assuming that these users are unfamiliar with any workflow, we can mute the effect of \textit{familiarity} by comparing users without prior experience that use the old wizard to the ones that use the redesigned wizard.\footnote{Note that we would also flag users as new users if they use a different browser or if they cleared their browser cache. However, this circumstance would only bias the results towards finding no difference between the new and the old wizard since we would incorrectly label experienced users as new users, making it less likely to find a decrease in the time it takes to create a HANA database instance with the new Provisioning Wizard. Hence, we are confident that, if anything, the true effect would be even stronger if we find an impact under these circumstances.} For the redesigned wizard, we do not restrict the sample to first time users since no user can be familiar with the new workflow.

Once we control for the effect of familiarity, we see that the differences between the two groups increase to $98.17$ seconds. On average, the successful creation of an instance in the new wizard is $36.3$\% faster than with the old one (95\% CI within [$21.79$\%; $48.12$\%]). Moreover, this difference is statistically significant (p-value = $2e^{-05}$), indicating that the new Provisioning Wizard indeed decreases the average time it takes for users to successfully create a HANA database instance once we control for the effect of familiarity.

Another relevant factor affecting the average time on task is the users' choice of how to configure the database. With the release of the redesigned wizard, users have several options: They can upload a JSON file, clone an existing configuration, use one of the preset options provided by the system, or choose the manual configuration as with the old wizard. Table \ref{tab:config} shows the frequency of each option based on all users who successfully completed the redesigned Provisioning Wizard:

\begin{table}
\caption{Frequency of Selected Configuration Options}\label{tab:config}
\begin{tabular}{l|cccc}
& Clone & JSON & Manual & Preset\\
HANA Database & 6 & 13 & 295 & 79
\end{tabular}
\end{table}

Since the manual configuration is essentially the same workflow as with the old wizard, we can compare the time users need to successfully create a database instance with the old wizard to the new wizard for all selection types except for the manual configuration. Overall, we compare $336$ operations from the old wizard to $393$ operations done with the new wizard. The average time with the old wizard is $193.15$ seconds. For the redesigned wizard, excluding cases with manual configuration selected, the average time is $159.75$ and the difference is statistically significant at the 10\% confidence level (p-value = $0.09$). Although the number of operations with the preset configuration is rather low ($79$ operations), a comparison between the old and the redesigned wizard using the preset reveals that the average time using the preset option is $143.29$ seconds, and the difference to the old wizard is statistically significant as well (p-value of $0.013$).

Finally, as additional sanity check, we can compare the old wizard to the new one with the manual configuration selection. Our rationale is that, since this is the only option available in the old Provisioning Wizard, we should not find a difference between the old and the new wizard if the user chooses the manual configuration option. In fact, the difference between these two groups is insignificant (p-value of $0.334$), providing additional confidence in the validity of our results.

To establish a causal relationship between the decrease in time on task and the redesigned Provisioning Wizard, we also implement the RD design discussed above. Again, we identify the MSE-optimal bandwidth using the empirical approach discussed earlier which identifies a window of 10 days before and after $\tau$ to be optimal for this use case \citep{calonico2015}. We estimate the LATE using a local log-linear regression model. Table \ref{tab:rdd_time} presents the results.

%  \caption{Coefficient Estimates from the Local Logistic Regression Model}\label{tab:rdd_bin} 

\begin{table} \centering 
  \caption{Coefficient Estimates from the Local Log-Linear Regression Model}
  \label{tab:rdd_time} 
\begin{tabular}{@{\extracolsep{5pt}}lc} 
\\[-1.8ex]\hline 
\hline \\[-1.8ex] 
 & \multicolumn{1}{c}{\textit{Dependent variable:}} \\ 
\cline{2-2} 
\\[-1.8ex] & log(time on task) – $log\left(Y^t\right)$ \\ 
\hline \\[-1.8ex] 
 T & $-$1.030$^{***}$ \\ 
  & (0.338) \\ 
  & \\ 
 R & 0.064 \\ 
  & (0.053) \\ 
  & \\ 
  T $\times$ R & 0.020 \\ 
  & (0.060) \\ 
  & \\ 
 Constant & 5.696$^{***}$ \\ 
  & (0.310) \\ 
  & \\ 
\hline \\[-1.8ex] 
Observations & 245 \\ 
R$^{2}$ & 0.066 \\ 
Adjusted R$^{2}$ & 0.054 \\ 
Residual Std. Error & 1.125 (df = 241) \\ 
F Statistic & 5.675$^{***}$ (df = 3; 241) \\ 
\hline 
\hline \\[-1.8ex] 
\textit{Note:}  & \multicolumn{1}{r}{$^{*}$p$<$0.1; $^{**}$p$<$0.05; $^{***}$p$<$0.01} \\ 
\end{tabular} 
\end{table}

Again, while the treatment indicator is statistically significant, we simulate the LATE to provide a more meaningful interpretation of the results. The simulation reveals that the redesigned wizard decreases the time by $98.41$ seconds [$29.02$ seconds; $164.24$ seconds] and that the decrease is statistically significant (empirical p-value = $0.003$). While showing some indication of a discontinuous change prior to the actual treatment assignment date $\tau$, the placebo cutoff approach described above generally confirms the validity of the RD approach.\footnote{The analysis finds statistically significant effects for the placebo cutoffs $\tau - 5$ (p-value = $0.01$) and $\tau - 7$ (p-value = $0.02$) but none of the other placebo cutoffs reveal statistically significant results at conventional significance levels. Once we account for multiple testing by applying the Bonferroni correction and adjust the significance level to $\alpha_{adj} = 0.008$, none of the placebo cutoffs exhibit statistically significant treatment effects.}

\section{Conclusion}
This study discusses the RD design as a viable research strategy to quantify the causal effect of UX changes on important business KPIs in the absence of a controlled experimental framework. Since causal statements remain essential for evidence-based product decisions, this approach is tailored to UX research in situations where cost, feasibility, or organizational constraints preclude randomized experiments. It provides a theoretical discussion of the method and the underlying assumptions as well as a mathematical derivation of the implications to obtain unbiased causal estimates. An empirical use case from the Provisioning Wizard in SAP HANA Cloud Central illustrates the practical applicability of the research strategy and its value for applied UX research in the software industry. By exploiting the natural threshold of the redesign deployment date, we identify local average treatment effects of the redesign on key metrics.

Based on usage tracking data from HANA Cloud Central, the analysis provides empirical evidence that the redesign of the Provisioning Wizard significantly increases the completion rate of HANA database instances. Using the quasi-experimental RD design, we are able to calculate local average treatment effects and establish a causal relationship between the new Provisioning Wizard and the completion rate. Regarding the time it takes users to successfully create a HANA database instance, we find no general impact since familiarity with the old process masks the efficiency gains the redesigned wizard offers. Once accounted for this, we find that, on average, users operating in the new wizard are $36.3$\% [$21.79$\%; $48.12$\%] faster in creating database instances than with the old one. The RD design further allows us to provide evidence for a causal relationship between the efficiency of the provisioning process and the new wizard.

Future research may explore other quasi-experimental techniques to causally evaluate the effect of UX improvements in business software products in the absence of a controlled experimental framework. Techniques like propensity score matching, difference-in-differences, instrumental variables, or synthetic control methods, for example, are frequently applied in other disciplines and have proven to be able to establish causality without a fully randomized treatment assignment. Exploring and utilizing these techniques will provide a strong empirical foundation for data-driven product improvements.

\bibliographystyle{authordate1}
\bibliography{ProvWiz_bibliography}

%%%%%%%%%%%%%%%%% APPENDICES %%%%%%%%%%%%%%%%%%%%%
\section*{Acknowledgement}
We thank Euree Hawer for her support.

%%%%%%%%%%%%%%%%%%%%%%%%%%%%%%%%%%%%%%%%%%%%%%%%%%

\label{lastpage}
\end{document}

%% file: distribution.tex
% Created by tikzDevice version 0.12.6 on 2026-03-19 13:24:13
% !TEX encoding = UTF-8 Unicode
\begin{tikzpicture}[x=1pt,y=1pt]
\definecolor{fillColor}{RGB}{255,255,255}
\path[use as bounding box,fill=fillColor,fill opacity=0.00] (0,0) rectangle (252.94,144.54);
\begin{scope}
\path[clip] (  0.00, 24.00) rectangle (102.47,132.54);
\definecolor{drawColor}{RGB}{0,0,0}
\definecolor{fillColor}{RGB}{211,211,211}

\path[draw=drawColor,line width= 0.4pt,line join=round,line cap=round,fill=fillColor] (  3.80, 28.02) rectangle ( 15.66,128.52);

\path[draw=drawColor,line width= 0.4pt,line join=round,line cap=round,fill=fillColor] ( 15.66, 28.02) rectangle ( 27.52, 33.56);

\path[draw=drawColor,line width= 0.4pt,line join=round,line cap=round,fill=fillColor] ( 27.52, 28.02) rectangle ( 39.38, 29.22);

\path[draw=drawColor,line width= 0.4pt,line join=round,line cap=round,fill=fillColor] ( 39.38, 28.02) rectangle ( 51.24, 29.07);

\path[draw=drawColor,line width= 0.4pt,line join=round,line cap=round,fill=fillColor] ( 51.24, 28.02) rectangle ( 63.10, 28.32);

\path[draw=drawColor,line width= 0.4pt,line join=round,line cap=round,fill=fillColor] ( 63.10, 28.02) rectangle ( 74.96, 28.32);

\path[draw=drawColor,line width= 0.4pt,line join=round,line cap=round,fill=fillColor] ( 74.96, 28.02) rectangle ( 86.82, 28.02);

\path[draw=drawColor,line width= 0.4pt,line join=round,line cap=round,fill=fillColor] ( 86.82, 28.02) rectangle ( 98.68, 28.32);
\end{scope}
\begin{scope}
\path[clip] (  0.00,  0.00) rectangle (252.94,144.54);
\definecolor{drawColor}{RGB}{0,0,0}

\node[text=drawColor,anchor=base,inner sep=0pt, outer sep=0pt, scale=  1.00] at ( 51.24,134.94) {\bfseries Original};

\path[draw=drawColor,line width= 0.4pt,line join=round,line cap=round] (  3.80, 24.00) -- ( 98.68, 24.00);

\path[draw=drawColor,line width= 0.4pt,line join=round,line cap=round] (  3.80, 24.00) -- (  3.80, 18.00);

\path[draw=drawColor,line width= 0.4pt,line join=round,line cap=round] ( 27.52, 24.00) -- ( 27.52, 18.00);

\path[draw=drawColor,line width= 0.4pt,line join=round,line cap=round] ( 51.24, 24.00) -- ( 51.24, 18.00);

\path[draw=drawColor,line width= 0.4pt,line join=round,line cap=round] ( 74.96, 24.00) -- ( 74.96, 18.00);

\path[draw=drawColor,line width= 0.4pt,line join=round,line cap=round] ( 98.68, 24.00) -- ( 98.68, 18.00);

\node[text=drawColor,anchor=base,inner sep=0pt, outer sep=0pt, scale=  0.80] at (  3.80, 10.80) {0};

\node[text=drawColor,anchor=base,inner sep=0pt, outer sep=0pt, scale=  0.80] at ( 27.52, 10.80) {2000};

\node[text=drawColor,anchor=base,inner sep=0pt, outer sep=0pt, scale=  0.80] at ( 51.24, 10.80) {4000};

\node[text=drawColor,anchor=base,inner sep=0pt, outer sep=0pt, scale=  0.80] at ( 74.96, 10.80) {6000};

\node[text=drawColor,anchor=base,inner sep=0pt, outer sep=0pt, scale=  0.80] at ( 98.68, 10.80) {8000};

\node[text=drawColor,anchor=base,inner sep=0pt, outer sep=0pt, scale=  1.00] at ( 51.24,  2.40) {Duration (sec)};
\end{scope}
\begin{scope}
\path[clip] (126.47, 24.00) rectangle (228.94,132.54);
\definecolor{drawColor}{RGB}{0,0,0}
\definecolor{fillColor}{RGB}{211,211,211}

\path[draw=drawColor,line width= 0.4pt,line join=round,line cap=round,fill=fillColor] (130.27, 28.02) rectangle (137.57, 30.32);

\path[draw=drawColor,line width= 0.4pt,line join=round,line cap=round,fill=fillColor] (137.57, 28.02) rectangle (144.86, 51.04);

\path[draw=drawColor,line width= 0.4pt,line join=round,line cap=round,fill=fillColor] (144.86, 28.02) rectangle (152.16, 86.33);

\path[draw=drawColor,line width= 0.4pt,line join=round,line cap=round,fill=fillColor] (152.16, 28.02) rectangle (159.46,100.90);

\path[draw=drawColor,line width= 0.4pt,line join=round,line cap=round,fill=fillColor] (159.46, 28.02) rectangle (166.76,123.15);

\path[draw=drawColor,line width= 0.4pt,line join=round,line cap=round,fill=fillColor] (166.76, 28.02) rectangle (174.06,128.52);

\path[draw=drawColor,line width= 0.4pt,line join=round,line cap=round,fill=fillColor] (174.06, 28.02) rectangle (181.36,115.48);

\path[draw=drawColor,line width= 0.4pt,line join=round,line cap=round,fill=fillColor] (181.36, 28.02) rectangle (188.66, 74.05);

\path[draw=drawColor,line width= 0.4pt,line join=round,line cap=round,fill=fillColor] (188.66, 28.02) rectangle (195.96, 63.31);

\path[draw=drawColor,line width= 0.4pt,line join=round,line cap=round,fill=fillColor] (195.96, 28.02) rectangle (203.25, 47.97);

\path[draw=drawColor,line width= 0.4pt,line join=round,line cap=round,fill=fillColor] (203.25, 28.02) rectangle (210.55, 36.46);

\path[draw=drawColor,line width= 0.4pt,line join=round,line cap=round,fill=fillColor] (210.55, 28.02) rectangle (217.85, 34.92);

\path[draw=drawColor,line width= 0.4pt,line join=round,line cap=round,fill=fillColor] (217.85, 28.02) rectangle (225.15, 31.09);
\end{scope}
\begin{scope}
\path[clip] (  0.00,  0.00) rectangle (252.94,144.54);
\definecolor{drawColor}{RGB}{0,0,0}

\node[text=drawColor,anchor=base,inner sep=0pt, outer sep=0pt, scale=  1.00] at (177.71,134.94) {\bfseries Log-Transformed};

\path[draw=drawColor,line width= 0.4pt,line join=round,line cap=round] (137.57, 24.00) -- (225.15, 24.00);

\path[draw=drawColor,line width= 0.4pt,line join=round,line cap=round] (137.57, 24.00) -- (137.57, 18.00);

\path[draw=drawColor,line width= 0.4pt,line join=round,line cap=round] (152.16, 24.00) -- (152.16, 18.00);

\path[draw=drawColor,line width= 0.4pt,line join=round,line cap=round] (166.76, 24.00) -- (166.76, 18.00);

\path[draw=drawColor,line width= 0.4pt,line join=round,line cap=round] (181.36, 24.00) -- (181.36, 18.00);

\path[draw=drawColor,line width= 0.4pt,line join=round,line cap=round] (195.96, 24.00) -- (195.96, 18.00);

\path[draw=drawColor,line width= 0.4pt,line join=round,line cap=round] (210.55, 24.00) -- (210.55, 18.00);

\path[draw=drawColor,line width= 0.4pt,line join=round,line cap=round] (225.15, 24.00) -- (225.15, 18.00);

\node[text=drawColor,anchor=base,inner sep=0pt, outer sep=0pt, scale=  0.80] at (137.57, 10.80) {3};

\node[text=drawColor,anchor=base,inner sep=0pt, outer sep=0pt, scale=  0.80] at (152.16, 10.80) {4};

\node[text=drawColor,anchor=base,inner sep=0pt, outer sep=0pt, scale=  0.80] at (166.76, 10.80) {5};

\node[text=drawColor,anchor=base,inner sep=0pt, outer sep=0pt, scale=  0.80] at (181.36, 10.80) {6};

\node[text=drawColor,anchor=base,inner sep=0pt, outer sep=0pt, scale=  0.80] at (195.96, 10.80) {7};

\node[text=drawColor,anchor=base,inner sep=0pt, outer sep=0pt, scale=  0.80] at (210.55, 10.80) {8};

\node[text=drawColor,anchor=base,inner sep=0pt, outer sep=0pt, scale=  0.80] at (225.15, 10.80) {9};

\node[text=drawColor,anchor=base,inner sep=0pt, outer sep=0pt, scale=  1.00] at (177.71,  2.40) {Natural Logarithm of Duration};
\end{scope}
\end{tikzpicture}

%% file: ProvWiz_bibliography.bib
@article{angrist1996,
author = {Joshua D. Angrist and Guido W. Imbens and Donald B. Rubin},
title = {Identification of Causal Effects Using Instrumental Variables},
journal = {Journal of the American Statistical Association},
volume = {91},
number = {434},
pages = {444–455},
year = {1996},
publisher = {Taylor \& Francis},
URL = {https://doi.org/10.2307/2291629}
}

@article{neyman1923,
 URL = {http://www.jstor.org/stable/2245382},
 author = {Jerzy Neyman},
 journal = {Roczniki Nauk Rolniczych Tom X {\rm [in Polish]; translated in} Statistical Science},
 number = {4},
 pages = {465–472},
 publisher = {Institute of Mathematical Statistics},
 title = {On the Application of Probability Theory to Agricultural Experiments. Essay on Principles. Section 9},
 urldate = {2026-03-17},
 volume = {5},
 year = {1923}
}

@article{rubin1974,
author = {Donald B. Rubin},
title = {Estimating causal effects of treatments in randomized and nonrandomized studies},
journal = {Journal of Educational Psychology},
volume = {66},
number = {5},
pages = {688–701},
year = {1974},
URL = {https://doi.org/10.1037/h0037350}
}

@article{rubin2005,
author = {Donald B. Rubin},
title = {Causal Inference Using Potential Outcomes},
journal = {Journal of the American Statistical Association},
volume = {100},
number = {469},
pages = {322–331},
year = {2005},
publisher = {Taylor \& Francis},
URL = {https://doi.org/10.1198/016214504000001880}
}

@article{holland1986,
author = {Paul W. Holland},
title = {Statistics and Causal Inference},
journal = {Journal of the American Statistical Association},
volume = {81},
number = {396},
pages = {945–960},
year = {1986},
publisher = {Taylor \& Francis},
URL = {https://doi.org/10.1080/01621459.1986.10478354}
}

@article{kohavi2009,
author = {Kohavi, Ron and Longbotham, Roger and Sommerfield, Dan and Henne, Randal M.},
title = {Controlled experiments on the web: survey and practical guide},
journal = {Data Mining and Knowledge Discovery},
volume = {18},
pages = {140–181},
year = {2009},
URL = {https://doi.org/10.1007/s10618-008-0114-1}
}

@article{kohavi2017,
author = {Kohavi, Ron and Thomke, Stefan},
title = {The Surprising Power of Online Experiments: Getting the Most Out of A/B and Other Controlled Tests},
journal = {Harvard Business Review},
volume = {95},
number = {5},
pages = {74–82},
year = {2017},
}

@article{quin2024,
title = {A/B testing: A systematic literature review},
journal = {Journal of Systems and Software},
volume = {211},
pages = {112011},
year = {2024},
issn = {0164-1212},
doi = {https://doi.org/10.1016/j.jss.2024.112011},
url = {https://www.sciencedirect.com/science/article/pii/S0164121224000542},
author = {Federico Quin and Danny Weyns and Matthias Galster and Camila Costa Silva},
}

@article{rubin1980,
 ISSN = {01621459, 1537274X},
 URL = {http://www.jstor.org/stable/2287653},
  title = {Randomization Analysis of Experimental Data: The Fisher Randomization Test Comment},
 author = {Donald B. Rubin},
 journal = {Journal of the American Statistical Association},
 number = {371},
 pages = {591–593},
 publisher = {[American Statistical Association, Taylor \& Francis, Ltd.]},
 volume = {75},
 year = {1980}
}

@article{larsen2024,
author = {Nicholas Larsen and Jonathan Stallrich and Srijan Sengupta and Alex Deng and Ron Kohavi and Nathaniel T. Stevens},
title = {Statistical Challenges in Online Controlled Experiments: A Review of A/B Testing Methodology},
journal = {The American Statistician},
volume = {78},
number = {2},
pages = {135–149},
year = {2024},
publisher = {Taylor \& Francis},
doi = {10.1080/00031305.2023.2257237},
URL = {https://doi.org/10.1080/00031305.2023.2257237}
}

@book{cox1958,
 author = {David R. Cox},
 publisher = {Wiley \& Sons},
 address = {New York},
 title = {Planning of Experiments},
 year = {1958}
}

@article{pearl2009,
author = {Judea Pearl},
title = {Causal inference in statistics: An overview},
volume = {3},
journal = {Statistics Surveys},
publisher = {Amer. Statist. Assoc., the Bernoulli Soc., the Inst. Math. Statist., and the Statist. Soc. Canada},
pages = {96–146},
year = {2009},
doi = {10.1214/09-SS057},
URL = {https://doi.org/10.1214/09-SS057}
}

@article{imbens2008,
title = {The regression discontinuity design—Theory and applications},
journal = {Journal of Econometrics},
volume = {142},
number = {2},
pages = {611–614},
year = {2008},
issn = {0304-4076},
doi = {https://doi.org/10.1016/j.jeconom.2007.05.008},
url = {https://www.sciencedirect.com/science/article/pii/S0304407607001200},
author = {Guido Imbens and Thomas Lemieux}
}

@article{lee2010,
author = {Lee, David S. and Lemieux, Thomas},
title = {Regression Discontinuity Designs in Economics},
journal = {Journal of Economic Literature},
volume = {48},
number = {2},
year = {2010},
pages = {281–355},
doi = {10.1257/jel.48.2.281},
URL = {https://www.aeaweb.org/articles?id=10.1257/jel.48.2.281}
}

@article{hahn2001,
 ISSN = {00129682, 14680262},
 URL = {http://www.jstor.org/stable/2692190},
 author = {Jinyong Hahn and Petra Todd and Wilbert Van der Klaauw},
 journal = {Econometrica},
 number = {1},
 pages = {201–209},
 publisher = {[Wiley, Econometric Society]},
 title = {Identification and Estimation of Treatment Effects with a Regression-Discontinuity Design},
 volume = {69},
 year = {2001}
}

@article{cattaneo2022,
   author = "Cattaneo, Matias D. and Titiunik, Roc\'{i}o",
   title = "Regression Discontinuity Designs", 
   journal= "Annual Review of Economics",
   year = "2022",
   volume = "14",
   pages = "821–851",
   doi = "https://doi.org/10.1146/annurev-economics-051520-021409",
   url = "https://www.annualreviews.org/content/journals/10.1146/annurev-economics-051520-021409",
   publisher = "Annual Reviews",
}

@article{thistlethwaite1960,
 URL = {https://doi.org/10.1037/h0044319},
 author = {Thistlewaite, Donald and Campbell, Donald},
 journal = {Journal of Educational Psychology},
 number = {6},
 pages = {309–317},
 title = {Regression-Discontinuity Analysis: An Alternative to the Ex-Post Facto Experiment},
 volume = {51},
 year = {1960}
}

@article{calonico2015,
  author = {Calonico, Sebastian and Cattaneo, Matias D. and Titiunik, Roc\'{i}o},
  title = {rdrobust: An R Package for Robust Nonparametric Inference in Regression-Discontinuity Designs},
  journal = {The R Journal},
  year = {2015},
  note = {https://doi.org/10.32614/RJ-2015-004},
  doi = {10.32614/RJ-2015-004},
  volume = {7},
  issue = {1},
  pages = {38–51}
}
